\documentclass{article}
\usepackage{tabulary}
\usepackage{color}
\usepackage[table]{xcolor}
\usepackage{multirow}
\usepackage{amsfonts,amsmath,amssymb}
\usepackage{graphicx}
\usepackage{caption}
\usepackage{tabulary}
\usepackage[]{siunitx}
\usepackage{authblk}
\usepackage[square,numbers]{natbib}
\usepackage{bm}
\usepackage{hyperref}
\hypersetup{
    colorlinks=true,
    linkcolor=blue,
    filecolor=magenta,      
    urlcolor=cyan,
}
\begin{document}

\title{A computational fluid dynamics model for the small-scale dynamics of wave, ice floe and interstitial grease ice interaction}

	\author[1]{Rutger Marquart}
	\author[2,3]{Alfred  Bogaers}
	\author[1]{Sebastian Skatulla}
	\author[4]{Alberto Alberello}
	\author[5]{Alessandro Toffoli}
    \author[6]{Carina Nisters}
	\author[7]{Marcello Vichi}

	\affil[1]{Computational Continuum Mechanics Research Group, Department of Civil Engineering, University of Cape Town, South Africa}
    \affil[2]{Ex Mente Technologies, Pretoria, South Africa}
    \affil[3]{School of Computer Science and Applied Mathematics, University of Witwatersrand, South Africa}
    \affil[4]{Graduate School of Frontier Sciences, The University of Tokyo, Japan}
    \affil[5]{Department of Infrastructure Engineering, University of Melbourne, Australia}
    \affil[6]{University of Duisburg-Essen, Institute of Mechanics, Germany}
    \affil[7]{University of Cape Town, Department of Oceanography, South Africa}

\maketitle

%	\begin{keyword}
%		Sea ice dynamics \sep Wave-ice interaction \sep Marginal ice zone \sep Sea ice rheology \sep Pancake ice \sep Grease ice.
%	\end{keyword}
	
	% Keywords
%\keyword{Sea ice dynamics; Wave-ice interaction; Marginal ice zone; Sea ice rheology; Pancake ice; Grease ice.} 

\abstract{%Large-scale phenomenological sea ice models have been developed, but
%for a wide range of applications. However, 
%ALESSANDRO - These first 2 sentences of the abstract do not introduce the problem we are addressing clearly. Why do we model collision dynamics?
%The highly dynamic sea ice behaviour in the marginal ice zone, the sea ice area affected by the wave motion, is not accurately captured by large-scale sea ice models due to their phenomenological nature fitting only the domain-averaged response.
%AT,AA -- 
The marginal ice zone is a highly dynamical region where sea ice and ocean waves interact. Large-scale sea ice models only compute domain-averaged responses. As the majority of the marginal ice zone consists of mobile ice floes surrounded by grease ice, finer-scale modelling is needed to resolve variations of its mechanical properties, wave-induced pressure gradients and drag forces acting on the ice floes. A novel computational fluid dynamics approach is presented, that considers the heterogeneous sea ice material composition and accounts for the wave-ice interaction dynamics. Results show, after comparing three realistic sea ice layouts with similar concentration and floe diameter, that the discrepancy between the domain-averaged temporal stress and strain rate evolutions increases for decreasing wave period. Furthermore, strain rate and viscosity are mostly affected by the variability of ice floe shape and diameter.
%, due to the heterogeneous morphology 
%In winter, the majority of the marginal ice zone consists of mobile ice floes surrounded by grease ice, therefore requiring detailed finer-scale modelling to resolve variations of its mechanical properties, wave-induced pressure gradients and drag forces acting on the ice floes. %sea ice dynamics models.
%This work introduces
%ALESSANDRO - From here on, the abstract sounds ok!
%Here we present
%A novel computational fluid dynamics approach is presented, % work describes key aspects of
%that considers the heterogeneous sea ice material composition and accounts for the wave-ice interaction dynamics. %, focusing on the mechanical properties of the sea ice rheology, such as viscosity, stress and strain rate. %, using a novel OpenFOAM sea ice model. 
%The model considers the heterogeneous sea ice material composition consisting of floes and grease ice in idealised and more realistic configurations, subjected to the action of typical water waves.
%Results show, after comparing three realistic sea ice layouts with similar concentration and floe diameter, that the discrepancy between the domain-averaged temporal stress and strain rate evolution increases for decreasing wave period. Furthermore, strain rate and viscosity are mostly affected by the change of the ice floe shapes. 
}

\section{Introduction} \label{Introduction}
% sea ice covers approximately $10 \%$ of the earth's surface and has a major impact on the global climate by modulating heat and momentum exchanges between the ocean and the atmosphere
%and life of marine organisms in and below the ice 
% \cite{Massom2010,Toppaladoddi2015,vichi2019effects}.
%. The ice, acting as a layer between the ocean and atmosphere, has a profound influence on the heat, gas and momentum exchange 
 %A large number of dynamic and thermodynamic factors make especially the Antarctic marginal ice zone (MIZ), 
The region where ocean processes affect sea ice, known as marginal ice zone (MIZ; \cite{wadhamsa2006wave}),
is a highly complex system \cite{Massom2010,alberello2020drift}
%requiring greater insight into the sea ice dynamics at the different stages of
particularly during the ice formation and melt season. %Understanding sea ice dynamics and thermodynamics results in better sea ice predictions and therefore a better comprehension of the MIZ \cite{Herman2016}. 
Most contemporary dynamic-thermodynamic sea ice models used for predicting global climate are phenomenological large-scale models (order of 100\,km \cite{hakkinen1987constitutive}) in which internal stresses are related to the strain rate via a viscous-plastic rheology \cite{hibler1979dynamic,keller1998gravity,wang2010}. %They have historically been developed for Arctic conditions, since the Arctic is more accessible than the Antarctic, as the Arctic is less remotely located and has less severe weather conditions.
%In these models, the viscous sea ice rheology relates the internal ice stress with the strain rate.
%Large sea ice areas with heterogeneous characteristics including fractures, leads and open water are modelled by means of a smeared model approach, in which several different types of ice are modelled as one homogeneous material.
%Large-scale sea ice models %, widely used in global climate modelling,
%and help to understand the impact of sea ice on seasonal changes. As these models use a smeared model approach, they parameterize the interaction between ice and ocean.
These models adopt a smeared approach in which an area with heterogeneous ice characteristics (fractures, leads, open water and ice type) is modelled as a single homogeneous material with averaged properties, and in which ice-ocean interactions are parameterised \cite{pritchard1988mathematical}.
The dynamics is governed by the momentum balance equation and two continuity equations controlling ice thickness and concentration, accounting for deformation and growth-related effects \cite{kleine1995mathematical}.
However, sea ice deformation cannot be exclusively described by a viscous-plastic rheology as sea ice drift on the scale of less than 10\,km is only accurate to a certain extent and fails to reproduce sea ice deformation at finer scales \cite{rampal2008scaling,Dansereau2016}.
%In \citet{rampal2008scaling} show that the viscous model disregards intermittency introduced by fractures as well as long-ranged elastic interactions and relaxation effects, in \citet{girard2011new} this was partly addressed by formulating the elastic stiffness in terms of a time scale-dependent damage parameter.
% The unders\citet{girard2011new}\citet{girard2011new}tanding of physical processes between the atmosphere and ocean in the Antarctic MIZ still requires attention due to the shortage of observations in the field. 
\\
\\
The MIZ is regulated by both large- and small-scale processes.
Large-scale processes, on a spatial scale of 100\,km and a temporal scale of several days, are the %e.g. the 
Coriolis force
%, caused by the earth's rotation, and sea surface dynamic height, which includes 
and large-scale ocean circulation \cite{mehlmann2017finite,mehlmann2017modified}. %, which in the Southern Ocean is dominated by mesoscale variability of the Antarctic circumpolar current.
Fine-scale processes, on the scale of hours and less than $10$ kilometers, are the variability in the sea ice cover and stress acting at the air-ocean-ice interface \cite{biddle2020observed,lepparanta2011drift}.
%Observations that have been done in the Antarctic MIZ are not sufficient to resolve the rapidly evolving dynamics of the ocean surface layer \cite{swart2020submesoscale}.
%Wave motion regulates pancake ice formation and growth, although in situ data to verify the existing theoretical sea ice models are very limited \cite{montiel2018attenuation,stopa2018strong}.
Among small-scale processes in the Antarctic MIZ, waves play a crucial role in the formation of sea ice \cite{shen2018remote,alberello2021extreme}. Furthermore, the propagation of waves from open ocean into the MIZ and wave-ice interaction are linked to wave scattering and dissipation through the momentum transfer to ice floes, which mainly depend on sea ice characteristics like ice floe geometry and floe size distribution \cite{toyota2016formation}.  In particular, at the ice edge, close to the open ocean, high energy incident waves create complicated sea states resulting in violent ice floe motion and collision \cite{squire2007ocean,shen2018remote}. 
During the Antarctic winter, when temperatures are low, sea ice initially forms as small circular floes \cite{shen1987role,alberello2019brief,alberello2019experimental} ($1-10\text{m}$ in diameter, smaller than the characteristic wavelength), known as pancakes, surrounded by grease ice under the action of ocean waves penetrating deep into the ice covered ocean \cite{kohout2014storm,stopa2018strong}.
\\
\\
Dynamic-thermodynamic sea ice models on a floe-scale are scarce \cite{Dansereau2016}, in particular with respect to the characteristic pancake ice floes found in the winter MIZ.
Most of existing fine-scale numerical models are based on a molecular dynamics schemes based on Hertzian collision dynamics \cite{shen1987role,hopkins1996mesoscale,Herman2016} that consider sea ice physics and dynamics at a floe level \cite{Herman2013}.
The Discrete-Element bonded-particle model by \citet{herman2019wave} highlights the significance of skin drag on wave attenuation and floe collision dynamics due to prolonged collision and reduced restitution coefficient.
The model uses an heuristic contact detection algorithm and includes simplified overwash (water overflowing the ice \cite{bennetts2015idealised,skene2015modelling,nelli2020water}), and elastic and inelastic contact force contributions, with the latter linked to the restitution coefficient; tangential friction is not accounted for. 
\citet{damsgaard2018application} uses a discrete element framework for the approximation of Lagrangian sea ice dynamics also at the floe level.
The modelling of fragmented sea ice in the MIZ is improved by using exact solutions for mechanical nonlinearities with arbitrary sea ice concentrations.
\citet{rabatel2015dynamics} describes the dynamics of an assembly of rigid ice floes using an event-driven algorithm \cite{mcnamara2011molecular}.
Their approach focuses on collisions of individual ice floes of both, arbitrary size and shape derived from satellite images from the Arctic.
Ice classified as ``new ice'' of recently frozen sea water that is not solid ice yet, such as grease ice, has not been considered in these models.
Moreover, with the exception of \citet{rabatel2015dynamics}, all other models have in common the limitation of using highly simplified ice floe shapes.
%The new-developed small-scale model takes into account these limitations, by considering realistically shaped pancake ice floes embedded in grease ice.
\\
\\
This paper introduces a novel computational framework, developed using the computational fluid dynamics (CFD) software OpenFOAM
%\footnote[1]{\url{https://www.openfoam.com/}}
, for the small-scale wave-ice interaction dynamics.
The governing equations for the small-scale model are introduced in Section~\ref{SSM} and the numerical implementation in Section~\ref{MS1}.
We focus on the pancake and grease ice rheology variables over short time periods (< 1 minute), during which thermodynamic effects on ice thickness and concentration are negligible and ice floe ridging can be disregarded.
We demonstrate the suitability of the proposed computational framework for modelling the elastic collision dynamics of free-floating ice floes embedded in interstitial grease ice under the action of ocean waves by providing a convergence analysis with regards to the domain grid refinement and the sea ice layout in Section~\ref{MS}.
We show that the dynamics depends on (i) ice characteristics, (ii) sea ice rheology and (iii) external forcing.
With these findings at hand, an initial investigation of realistic ice floe distributions is undertaken in Section~\ref{SA}, focusing on the inter-dependency of floe size, wave forcing and grease ice viscosity.
%This study investigates the influence of the sea ice layout composition on the grease ice and pancake ice floe rheology variables for three different wave characteristics.
Final remarks and conclusions are presented in Section~\ref{D}.
%%%%%%%%%%%%%%%%%%%%%%%%%%%%%%%%%%%%%%%%%%

\section{Small-scale model}\label{SSM}

\subsection{Momentum equation}

\noindent
The sea ice dynamics is governed by the momentum balance equation
\begin{align}
m \left( \dfrac{\partial \bm{U}}{\partial t} + (\bm{U} \cdot \bm{\nabla}) \bm{U}\right) = \bm{\tau}_a + \bm{\tau}_o + \bm{\tau}_{w} + \bm{\nabla} \cdotp \bm{\sigma}, \label{eq:198}
\end{align}
where $\bm{U}$ represents the sea ice velocity vector and $m$ the mass of ice per area
\begin{align}
m = \rho_i h, \label{eq:mass}
\end{align}
where $\rho_i$ is the ice density and $h$ the ice thickness.
The internal reaction forces are represented by the \textit{Cauchy} stress tensor, $\bm{\sigma}=\{\bm{\sigma}_{grease},\bm{\sigma}_{floe}\}$, depending on the ice constituent.
In-plane wind and ocean current stress vectors applied to the ice, i.e.~the external forcing, are represented by $\bm{\tau}_a$ and $\bm{\tau}_o$, respectively.
The in-plane stress due to the waves, $\bm{\tau}_{w}$ (also an external forcing), is derived from the linear wave theory \cite{holthuijsen2010waves} and consists of two components:
\begin{align}
\bm{\tau}_{w}  &= \bm{\tau}_{sd} + \bm{\tau}_{fk},
\label{eq:1111}
\end{align}
where $\bm{\tau}_{sd}$ is the viscous component representing the skin drag acting on the ice floe basal plane and $\bm{\tau}_{fk}$ the Froude-Krylov force due to wave pressure field acting on the ice floe circumference.
\\
\\
The skin drag is given as
\begin{align}
\bm{\tau}_{sd}  &= -\rho_w C_w \vert \bm{U}_{w} - \bm{U} \vert \cos \theta_w \bm{U} +  \rho_w C_w \vert \bm{U}_{w} - \bm{U} \vert (\bm{U}_{w} \cos \theta_w + ((\bm{U}_{w} - \bm{U}) \times \bm{k}) \sin \theta_w),
\label{eq:skin drag}
\end{align}
where $\rho_w$ and $\theta_w$ represent the water density and the ice-ocean turning angle, respectively, $\bm{k}$ a unit normal vector to the surface of the ice and the ice-ocean drag coefficient, $C_w$, is composed of different values for pancake ice floes and grease ice, which can be found in Table~\ref{tab1}.
The orbital velocity of the water, $\bm{U}_{w}$, for monochromatic waves is defined as
\begin{align}
\bm{U}_{w} = \begin{pmatrix} U_{w_x} \\ U_{w_y} \\ U_{w_z} \end{pmatrix} = \begin{pmatrix} a\omega \sin(\omega t - k x) \\ 0 \\  a\omega \cos(\omega t - k x) \end{pmatrix},
\label{eq:Uwa}
\end{align}
where the \textit{x} is taken along the main direction of wave propagation and \textit{z} in the vertical direction, such that the \textit{y}-velocity component can be assumed zero. In Equations~\eqref{eq:Uwa} and \eqref{eq: wave-induced pressure} $a$, $\omega$ and $k$ represent the wave amplitude, wave frequency and wave number, respectively. The wave frequency and wave number, $\omega=2\pi/T$ and $k=2\pi/\lambda$, are computed from the wave period $T$ and the wavelength $\lambda$ using the deep water dispersion relation $\omega^2=gk$.
\\
\\
The Froude-Krylov force, $\bm{\tau}_{fk}$, accounts for the horizontal surge force due to the wave-induced pressure acting at the interface between ice floe and water \cite{herman2018wave}
\begin{equation}
\bm{\tau}_{fk} = -\int_{h_w} p \bm{n} \text{d} h_w, \label{eq: horizontal surge force}
\end{equation}
where $h_w$ represents the submerged ice floe thickness portion, calculated assuming that $90\%$ of both ice types is submerged.
%\textcolor{red}{WHERE DO 0.8 and 0.2 COME OUT FROM? can probably be rmoved this simple equation}
%\begin{equation}
%0.9(h_{floe}-h_{grease}) = 0.9(0.8-0.2) = 0.54 \text{m}.    
%\end{equation}
The unit vector $\bm{n}$ acts normal to the circumference of the ice floes directed outwards. The wave-induced pressure, $p$, is written as
\begin{equation}
p = \rho_w g a \sin(\omega t - k x), \label{eq: wave-induced pressure}
\end{equation}
where the water density and gravitational acceleration are denoted as $\rho_w$ and $g$, respectively.
%, which is used to define the wave steepness as $ka$. 
\\
\\
Equations~\eqref{eq: horizontal surge force} and \eqref{eq: wave-induced pressure} can be re-written as \cite{herman2018wave}
\begin{equation}
\bm{\tau}_{fk} = \bm{n} \cdot h_w a\omega^2 \cos(\omega t - k x). \label{eq: horizontal surge force version 2}
\end{equation}
The form drag acting on the ice floe circumference due velocity differences of floes and surrounding grease ice is implicitly included by the continuum model comprising both ice constituents enforcing velocity continuity.
\\
\\
Ocean current drag, $\bm{\tau}_o$, at the basal plane of the ice due to relative velocity of the ocean current and the ice is given by
\begin{align}
\bm{\tau}_o \ & = -\rho_w C_w \vert \bm{U}_o - \bm{U} \vert \cos \theta_w \bm{U} +  \rho_w C_w \vert \bm{U}_o - \bm{U} \vert (\bm{U}_o \cos \theta_w + ((\bm{U}_o - \bm{U}) \times \bm{k}) \sin \theta_w),
\label{eq: basal drag}
\end{align}
whereas the wind drag, $\bm{\tau}_a$, on the apical plane by
\begin{align}
\bm{\tau}_a = \rho_a C_a \vert \bm{U}_a \vert \left(\bm{U}_a \cos \theta_a + \bm{k} \times \bm{U}_a \sin \theta_a \right)\,.
\label{eq: tau_a}
\end{align}
The variables $\bm{U}_a$ and $\bm{U}_o$ represent the velocities of wind and ocean boundary layers, respectively. The air density, ice-air drag coefficient, and wind turning angle are indicated by $\rho_a$, $C_a$ and $\theta_a$, respectively.

\subsection{Sea ice rheology}

The rheology in the model is based on the material characteristics of each constituent, grease ice and pancake floes, respectively, and each is represented by its own rheology.
Grease ice behaves like a fluid whereas ice floes have a solid-like behaviour.
The ice thickness of both is assumed spatially constant, since ice thickness is negligibly small compared to the domain size in lateral direction.
As mentioned before, since only small time periods are considered in the model, ice thickness is assumed unaffected by thermodynamic and rafting-related growth and does not vary in time.
%Two ice constituents are distinguished, i.e. grease ice and ice floes, where both are represented by their own rheology, given below.
\\
\\
The sea ice strain rate tensor, $\dot{\bm{\epsilon}}$, is used in both grease ice and ice floe rheology, and is written in terms of the sea ice velocity vector, $\bm{U}$, as
\begin{align}
\dot{\bm{\epsilon}} = \dfrac{1}{2}(\bm{\nabla} \bm{U} + (\bm{\nabla} \bm{U})^T).\label{eq:6} 
\end{align}

\subsubsection{Grease ice rheology}

The fluid-like viscous-plastic behaviour of grease ice is modelled applying a viscous-plastic flow rheology similar as proposed by e.g. Thorndike and Rothrock \cite{Thorndike1975} or Hibler \cite{hibler1979dynamic}
%, which relates the sea ice stress tensor components, $\bm{\sigma}$, and the sea ice strain rate tensor components, $\dot{\bm{\epsilon}}$
\begin{align}
{\bm{\sigma}_{grease}} = 2\eta\dot{\bm{\epsilon}} + \bm{I} \left( (\zeta-\eta) \text{tr}(\dot{\bm{\epsilon}})-\dfrac{P}{2} \right) \label{eq:51},
\end{align}
where $P$ represents the internal grease ice strength that controls its compressibility, $\zeta$ and $\eta$ are strain rate-dependent grease ice viscosities, namely the spherical and deviatoric contributions, respectively. 
The main difference to the above mentioned models is the definition of the ice strength parameter as
\begin{align}
P = P^{*}h \label{eq:150}
\end{align}
where $P^{*}$ represents an empirical constant excluding ice growth and concentration effects. Both strain rate-dependent viscosities are coupled via
\begin{align}
\zeta &= \dfrac{P}{2\Delta} \quad\text{and}\quad \eta = \dfrac{\zeta}{e^2}, \label{eq:8} 
\end{align}
where the ratio between the in-plane principle axes of the elliptical yield curve is given by parameter, $e$, and the effective strain rate, $\Delta$, is given by
\begin{align}
\Delta = \left( (\dot{\epsilon}_{11}^2 +\dot{\epsilon}_{22}^2)(1+e^{-2}) + 4e^{-2}\dot{\epsilon}_{12}^2 + 2\dot{\epsilon}_{11}\dot{\epsilon}_{22}(1-e^{-2}) \right)^{1/2}. \label{eq:9999}
\end{align}
The Cartesian components of the strain rate tensor are denoted by $\dot{\epsilon}_{11}$, $\dot{\epsilon}_{22}$ and $\dot{\epsilon}_{12}$. 
As the strain rate approaches zero, the viscosity tends to infinity. To address this, a lower limit of the effective strain rate is enforced, i.e. $\Delta = 2 \cdot 10^{-7} \text{s}^{-1}$ \cite{Lepparanta1985}.
\\
\\
Equation~\eqref{eq:51} can be, after substitution of Equation~\eqref{eq:6}, written in terms of the velocity vector, $\bm{U}$, as 
\begin{align}
\bm{\sigma}_{grease} = \eta(\bm{\nabla} \bm{U} + (\bm{\nabla} \bm{U})^T)  + \bm{I} \left( (\zeta - \eta) \text{tr}(\bm{\nabla} \bm{U}) - \dfrac{P}{2} \right). \label{eq:1313}
\end{align}

\subsubsection{Ice floe rheology}

The solid-like material response of ice floes exhibits relatively small deformations and its constitutive behaviour floes is therefore described using the generalised Hooke's law
\begin{align}
\bm{\sigma}_{floe} & = 2 \mu \bm{\epsilon} + \lambda \bm{I} \text{tr}(\bm{\epsilon}), \label{eq:13}
\end{align}
where $\mu$ and $\lambda$ represent the $\text{Lam\'{e}}$ constants. Note as \textit{Hooke's} law is a linear function of the strain,  $\bm{\epsilon}$, %In order to write \textit{Hooke's} law in terms of the strain rate, its linearisation in time is required. Accordingly, 
Equation~\eqref{eq:6} is substituted in the linearised equation to write it in terms of the velocity vector, $\bm{U}$, resulting in 
\begin{align}
\bm{\sigma}^{n+1}_{floe} = \bm{\sigma}^n + \Delta t \left(\mu (\bm{\nabla} \bm{U} + (\bm{\nabla} \bm{U})^T) + \lambda \bm{I} \text{tr}(\bm{\nabla} \bm{U}) \right), \label{eq:linearised stress}
\end{align}
where $n$ represents the time index discretised with time step $\Delta t$ \cite{Lee1988}.

\section{Numerical implementation}\label{MS1}

The framework described in Section~\ref{SSM} is implemented in the Computational Fluid Dynamics (CFD) software OpenFOAM, which is based on the finite volume method (FVM) where a continuum body is approximated by a discrete model \cite{Ferziger2003}.
The domain is subdivided, by means of a mesh, into a finite number of small control volumes (cells). The volume of fluid (VOF) method is used in OpenFOAM to distinguish between two immiscible and incompressible fluids based on the non-dimensional parameter $\alpha$ \cite{roenby2016computational, roenby2019isoadvector}. %The VOF method is a numerical technique, describing the interface between two immiscible and incompressible fluids \cite{roenby2016computational,roenby2019isoadvector}.
%This method determines volume fractions of fluids across the cell faces per time step.
The VOF method is complemented with an interface compression scheme \cite{roenby2016computational} that enforces a sharp interface between dominantly solid-like pancake ice floes and viscous fluid-like interstitial grease ice.
Cells containing exclusively pancake ice or grease ice have $\alpha$ values of 1 and 0, respectively, and all intermediate values ($0<\alpha<1$) have no physical interpretation but are needed in the FVM which does not allow for discontinuities.
The region of $0<\alpha<1$ should be numerically constrained to the thin interface between the two ice materials. We note that as both ice floes and grease ice are modelled in a continuum fashion, the floe-floe and the floe-grease ice interactions are naturally accounted for in terms of their normal and tangential force components.
\\
\\
Ice floes and interstitial grease ice are discretised in OpenFOAM with finite volume elements using the least-squares gradient scheme and the \textit{Gauss} linear divergence scheme for the spatial integration. The \textit{Euler} implicit method is used for temporal discretisation. 
\\
\\
The explicit influence of the grease ice viscosity on wave attenuation is commonly not accounted for, because dissipation per wavelength is small ($<0.5\%$) \cite{alberello2021extreme}, i.e.~ the wave amplitude remains almost constant within a small domain. The domain boundary is numerically approximated by horizontal zero-gradient boundary conditions. As the wave forcing is imposed, the boundary condition introduces an error but only to cells directly adjacent to the boundary itself but does not affect the velocity field in the inner domain. To overcome this inconsistency the domain is enlarged to exclude undesired boundary effects for the inner domain.
%The zero-gradient boundary conditions are used in all simulations performed in this paper.
All terms in the momentum balance equation, Equation~\eqref{eq:198}, are normalized by the respective height of pancake and grease ice, resulting in stress values in terms of $\text{kg s}^{-2}$. Table~\ref{tab1} shows the used simulation parameters. The ice-ocean drag coefficients are based on values found in \cite{lu2011parameterization, mcguinness2009frazil, andreas1984atmospheric}. 
\\
\\
The convergence analysis with regard to grid size and domain size is conducted to find the optimal grid size to be used in the discretisation of the FVM (in Section~\ref{GCA}) and the critical ratio of floe diameter to domain size as linked to the imposed wave forcing (in Section~\ref{DCA}).
Simplified ice layouts are used consisting of randomly distributed disk-shaped pancake ice floes of constant diameter surrounded by grease ice with a temporally and spatially varying viscosity $\nu \approx 0.04 \text{m}^2 \text{s}^{-1}$. In the grid size convergence analysis the domain is exposed to a harmonic propagating wave with period $T = 18 \text{s}$ ($\lambda = 506 \text{m}$) and amplitude $a = 4.8 \text{m}$, typically encountered in the Antarctic MIZ \cite{alberello2020drift}. In the domain size convergence analysis three different wave forcing scenarios are considered with periods $T = 6 \text{s}$, $T = 12 \text{s}$ and $T = 20 \text{s}$ (corresponding to $\lambda = 56 \text{m}$, $\lambda = 225 \text{m}$ and $\lambda = 625 \text{m}$) to cover the majority of occurring wave conditions and amplitudes $a = 0.5 \text{m}$, $a = 2.1 \text{m}$ and $a = 6 \text{m}$, respectively.
The wave parameters are chosen to maintain a constant wave steepness ($ak$) across the considered cases corresponding to storm waves propagating into the MIZ to highlight their effect on heterogeneous sea ice conditions \cite{alberello2021extreme,alberello2021ijope}. The time step $\Delta t=0.01\text{s}$ for the Euler implicit method provides numerically stable simulations.
\\
\\
To study the high-resolution mechanical response of sea ice due to wave-ice interaction and ice composition in terms of ice type as well as floe shape and diameter in Section~\ref{SA}, three realistic  $100 \times 100 \text{m}^2$-sea ice layouts are extracted from in-situ image and video material recorded by stereo cameras of the dynamics of sea ice in the Antarctic MIZ \cite{alberello2019brief}. The images are collected in close vicinity to each other and, therefore, in homogeneous sea ice conditions with only slightly differing sea ice properties in terms of ice floe concentration and %$57.9\%$ ($\pm 3.2\%$) and % with a margin of approximately $5\%$. 
median ice floe caliper diameter.
%with $10.8\text{m}$ ($\pm 2.2\text{m}$) in \textit{x}-direction and $8.7\text{m}$ ($\pm 1.7\text{m}$) in \textit{y}-direction, respectively. %with a margin of $10\%$.
The different layouts also help to verify the robustness of the numerical framework with respect to natural variability of the ice floe layout in the MIZ.

\begin{table}[h] 
\captionsetup{justification=centering}
\begin{center}
\caption{General parameters used in all simulations.\label{tab1}}
\begin{tabular}{l l l l}
%\multicolumn{4}{c}{\textbf{General parameters used in all analysis}}\\
\hline
\textbf{Parameter} & \textbf{Definition} & \textbf{Value} & \textbf{Unit} \\
\hline
 $h_{i,g}$ & thickness ice floes and grease ice  &  0.8, 0.2 & $\text{m}$  \\
  $h_{w}$ & submerged ice floe thickness & 0.54 & $\text{m}$  \\
 $\rho_{i,g}$ & density ice floes and grease ice  &  918, 930 & $\text{kg\,m}^{-3}$ \\
$\lambda_i$ & first Lam\'{e} parameter ice floes & 6.4e6 & $\text{N\,m}^{-2}$ \\
$\mu_i$ & second Lam\'{e} parameter ice floes & 3.3e6 & $\text{N\,m}^{-2}$ \\
   $\rho_{w}$ & water density & 1026 & $\text{kg\,m}^{-3}$ \\
  $C_{w,i,g}$ & ice-ocean drag coefficient ice floes and grease ice & 0.005, 0.0013 & -\\
  $\theta_{w}$ & ice-ocean turning angle & 0 & $^{\circ}$\\
  $e$ & yield surface axes ratio grease ice& 2 & - \\ 
  $ak$ & wave steepness & 0.06 & - \\
  \hline
\end{tabular}
\centering
\label{tab:1}
\end{center}
\end{table}
%%%%%%%%%%%%%%%%%%%%%%%%%%%%%%%%%%%%%%%%%%

\section{Convergence analysis}\label{MS}

\subsection{Grid size convergence analysis} \label{GCA}

\begin{figure}[h]
\begin{center}
\includegraphics[width=9cm]{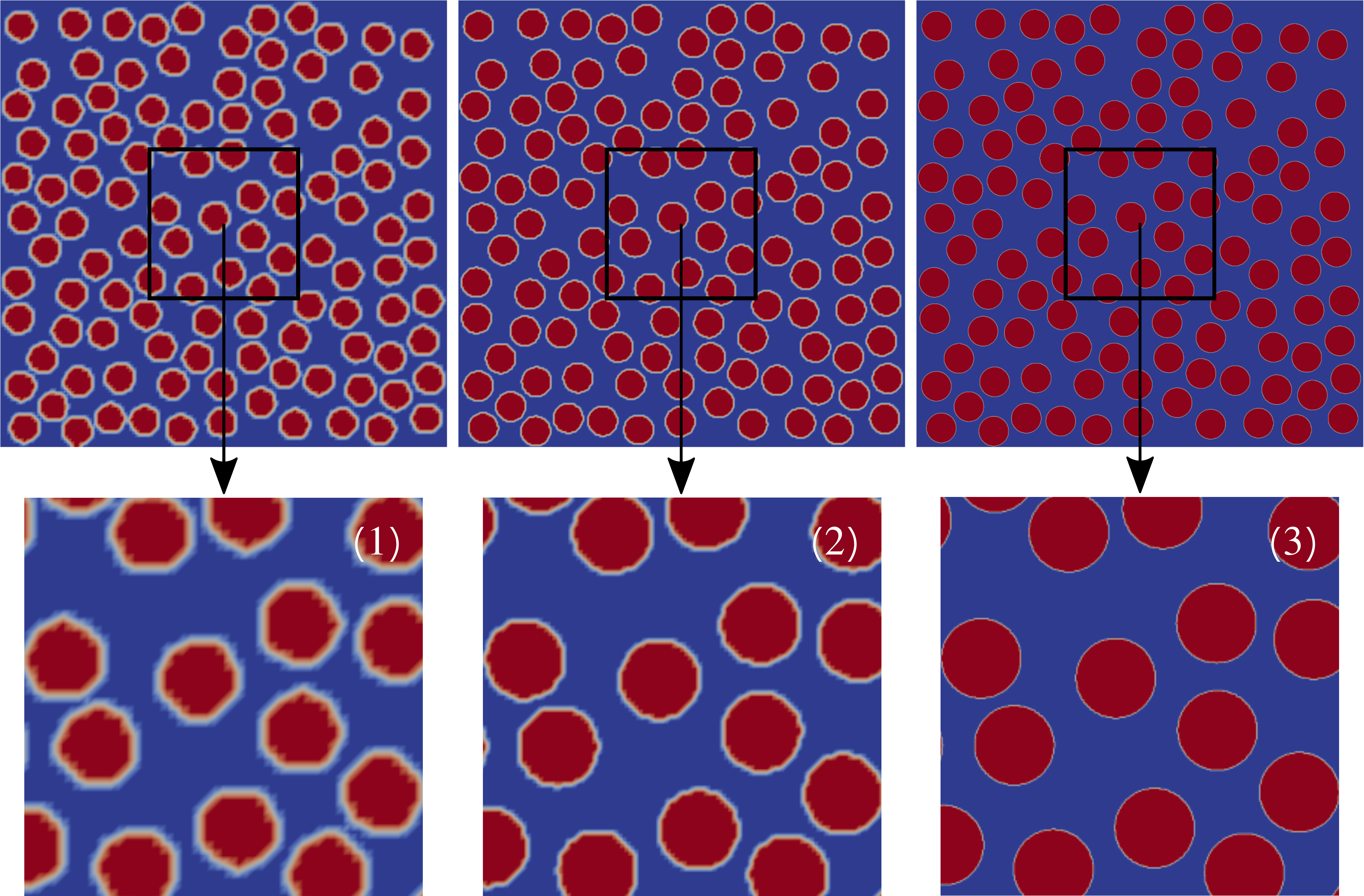}
\caption{Three different grid refinement levels of a $300 \times 300 \text{m}^2$-outer domain of randomly distributed disk-shaped $20 \text{m}$- diameter pancake floes each with $100 \times 100 \text{m}^2$-inner domains of following discretisations: (1) 2500 cells with a $2 \times 2 \text{m}^2$ cell size, (2) 10,000 cells with a $1 \times 1 \text{m}^2$ cell size and (3) 160,000 cells with a $0.25 \times 0.25 \text{m}^2$ cell size. Red and blue represent ice floes and grease ice, respectively. The numerical interface is indicated by the white colour. \label{fig1}}
\end{center}
\end{figure}

To accurately resolve the interface of pancake floes and the surrounding grease ice, discretisations featuring a higher number of cells provide a sharper boundary as indicated by the white colour in the pancake-grease ice layouts depicted in Figure~\ref{fig1} at $t=0 \text{s}$. For each discretisation refinement level, the domain-averaged stress and strain rate magnitudes, $\sigma_{\text{mag}}$ and $\dot{\epsilon}_{\text{mag}}$, as well as the bulk viscosity, $\zeta$, are computed for comparison.
%Once the average is unaffected by changes in cell size, a suitable mesh is deemed to be found.  
\\
\\
A $100 \times 100 \text{m}^2$-inner domain size with a $41\%$ ice floe area fraction is embedded in a $300 \times 300 \text{m}^2$ outer domain to exclude undesired boundary effects for the inner domain, as shown in Figure~\ref{fig1} by the black rectangle. The ice floes have a constant diameter of $20 \text{m}$. Multiple simulations are carried out, each time with an increasing number of cells in the domain ranging from 2500 to 160,000 cells.

\begin{figure}[h]
\includegraphics[width=13.5cm]{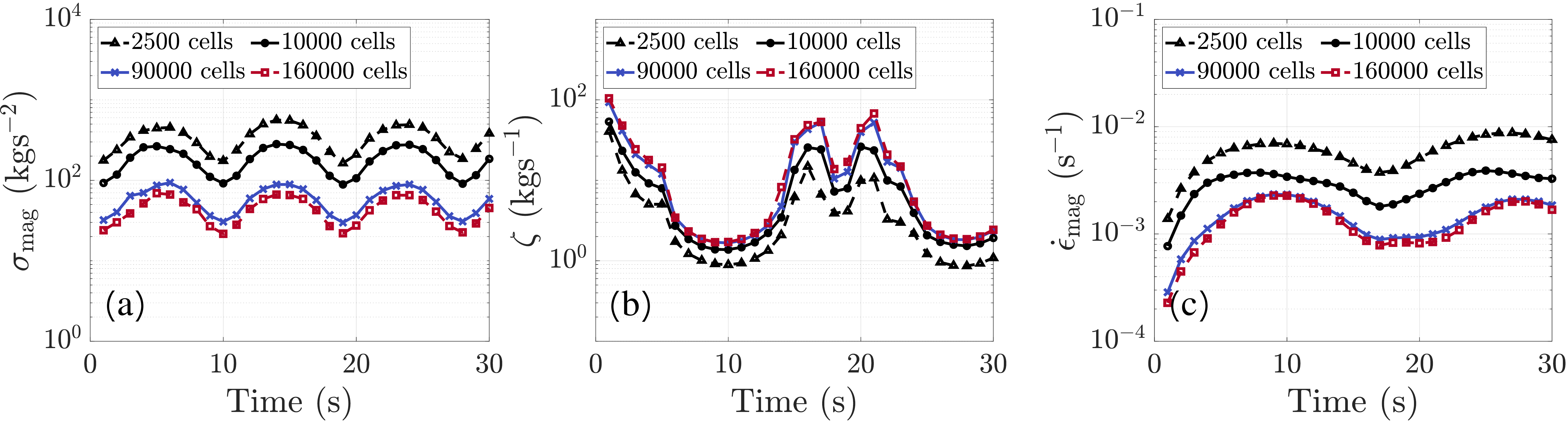}
\caption{Comparison of the spatial average of the main sea ice rheology variables for four different discretisation sizes: \textbf{(a)} stress magnitude, \textbf{(b)} bulk viscosity and \textbf{(c)} strain rate magnitude.\label{fig2}}
\end{figure}

%The stress and strain rate magnitude are calculated according to Equations~\eqref{eq:1115} and \eqref{eq:15}, respectively:
%
%\begin{equation}
%\sigma_{\text{mag}} = \sqrt{\sigma_{ij} \sigma_{ij}}\label{eq:1115} 
%\end{equation}
%\begin{equation}
%\dot{\epsilon}_{\text{mag}} = \sqrt{\dot{\epsilon}_{ij} \dot{\epsilon}_{ij}} \label{eq:15}
%\end{equation}
%
%where $\sigma_{ij}$ and $\dot{\epsilon}_{ij}$ represent the Cartesian components of both the stress and strain rate tensor, respectively. 
Figure~\ref{fig2} shows the domain-averaged stress, viscosity and strain rate results for the four different grid refinement levels within the $100 \times 100 \text{m}^2$-inner domain. The blue and red curves, representing 90,000 and 160,000-cell-inner domains, respectively, show similar values indicating convergence. It can be concluded that a total of 90,000 cells for the inner domain, corresponding to a grid size of $0.33 \times 0.33 \text{m}^2$, is sufficient and will be used for all numerical simulations in the subsequent sections.

\subsection{Domain size convergence analysis} \label{DCA}

\begin{figure}[h]
\begin{center}
\includegraphics[width=5cm]{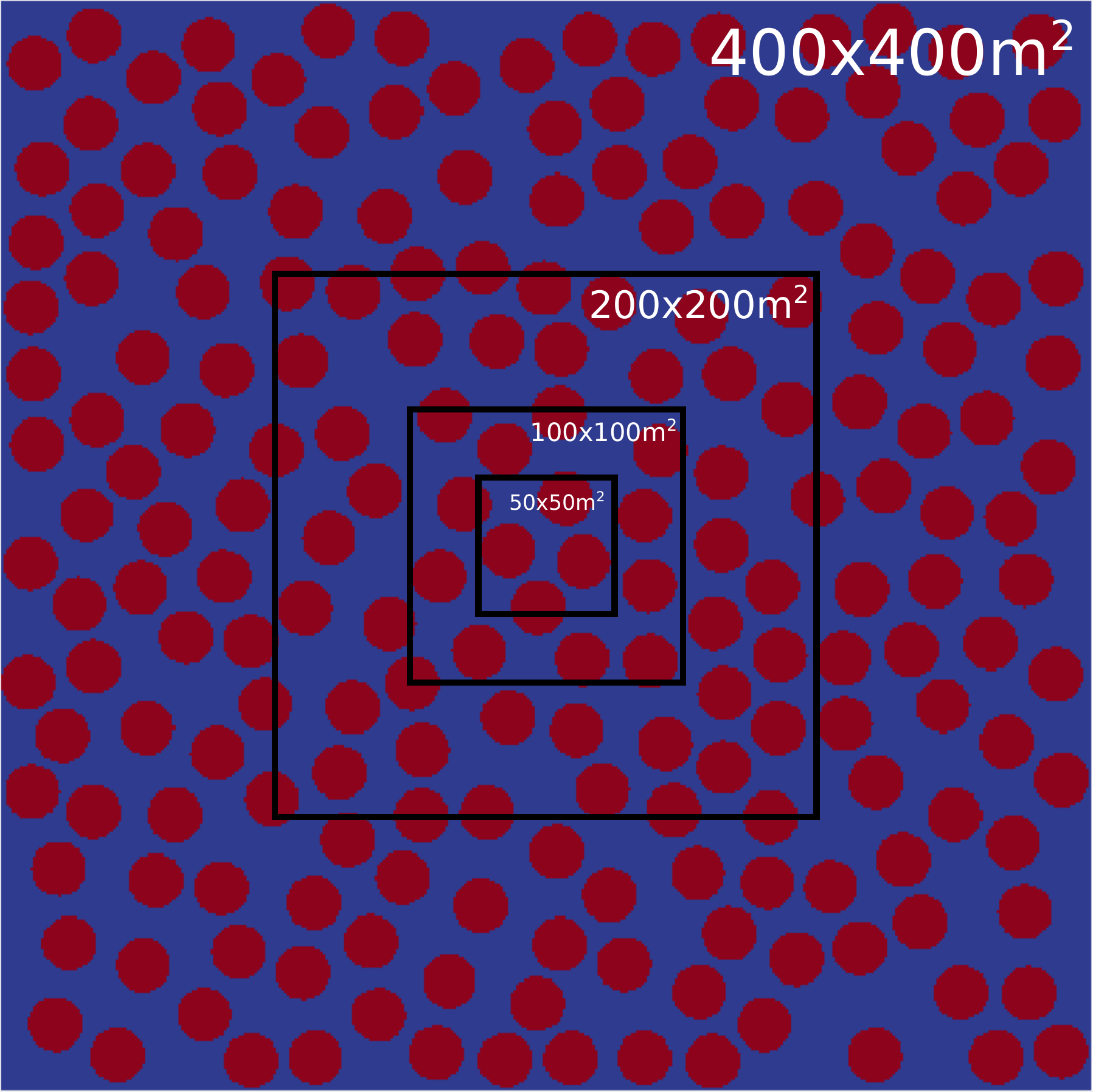}
\caption{Inner domain sizes ranging from $50 \times 50 \text{m}^2$ to $400 \times 400 \text{m}^2$ with varying ice floe concentration. Ice floes are modelled as disk-shaped floes of $20 \text{m}$-diameter. Red and blue represent ice floes and grease ice, respectively. \label{fig3}}
\end{center}
\end{figure}

The threshold for convergence with respect to the minimum required domain size is found by calculating the domain-averaged stress and strain magnitudes, $\sigma_{\text{mag}}$ and $\dot{\epsilon}_{\text{mag}}$, as well as bulk viscosity, $\zeta$, for increasing inner domain sizes. Once the average is unaffected by changes in the domain size, the transition from small- to large-scale modelling is identified, and the link to a phenomenological model using homogeneous material properties can be established.
%This allows us to study the detailed mechanical response of the sea ice rheology on smaller scale, which is different for any domain size smaller than the threshold.
\\
\\
A total domain size of $3600 \times 3600 \text{m}^2$ is considered, from which results are extracted for smaller inner domains, ranging from $50 \times 50 \text{m}^2$ to $3200 \times 3200 \text{m}^2$. This ensures a boundary zone of $400 \text{m}$ around the largest inner domain to exclude unwanted boundary effects due to the zero-gradient boundary conditions. Figure~\ref{fig3} shows the smallest four inner domain sizes ranging from $50 \times 50 \text{m}^2$ to $400 \times 400 \text{m}^2$.

\begin{figure}[h]
\includegraphics[width=13.5cm]{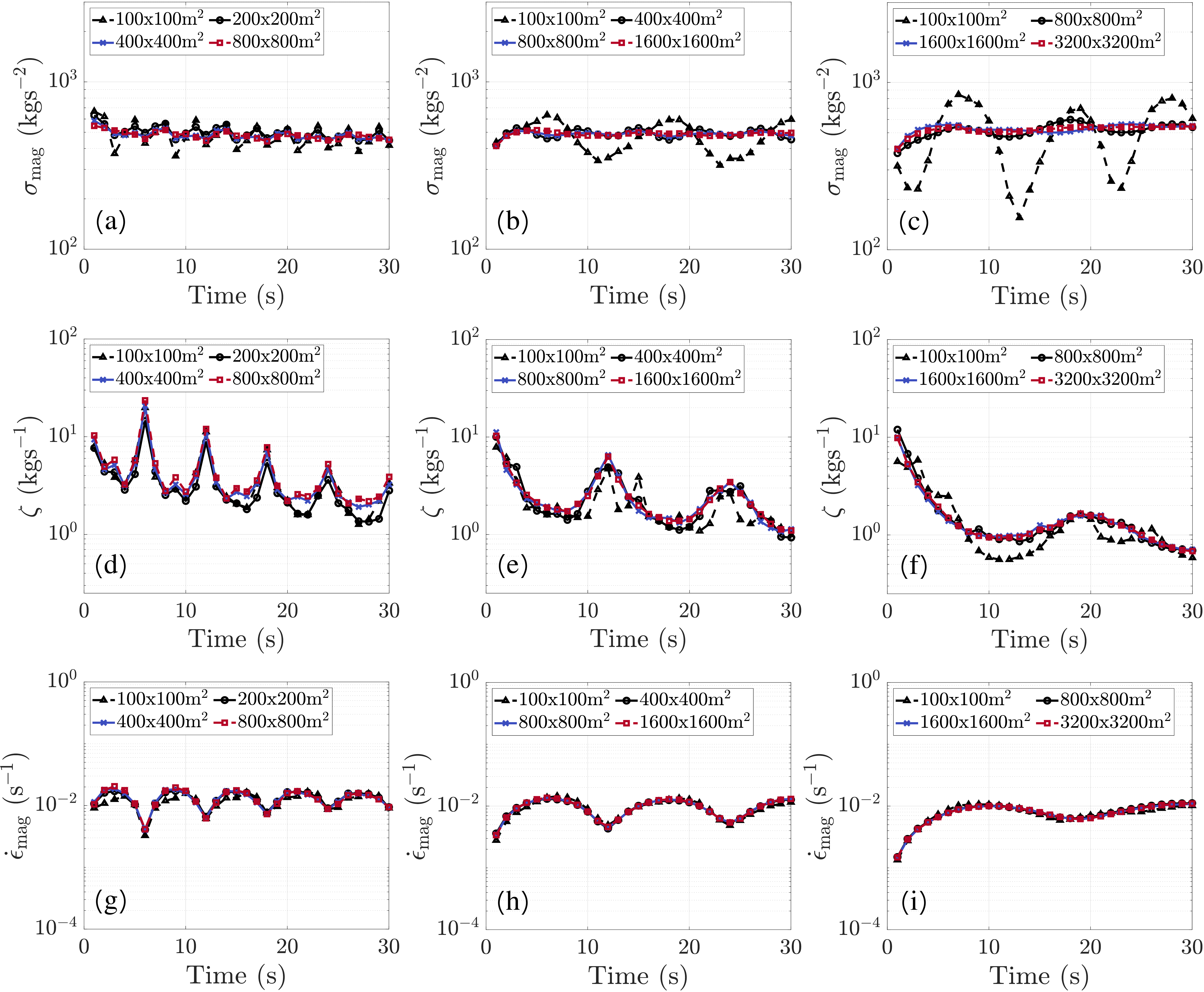}
\caption{Comparing the domain-averaged  sea ice rheology variables for different inner domain sizes and three different wave periods. Stress magnitude for $T = 6 \text{s}$ \textbf{(a)}; for $T = 12 \text{s}$ \textbf{(b)}; for $T = 20 \text{s}$ \textbf{(c)}. Bulk viscosity for $T = 6 \text{s}$ \textbf{(d)}; for $T = 12 \text{s}$ \textbf{(e)}; for $T = 20 \text{s}$ \textbf{(f)}. Strain rate magnitude for $T = 6 \text{s}$ \textbf{(g)}; for $T = 12 \text{s}$ \textbf{(h)}; for $T = 20 \text{s}$ \textbf{(i)}. \label{fig4}}
\end{figure}

With $20 \text{m}$-diameter ice floes the ice floe concentration from the largest domain size, $3600 \times 3600 \text{m}^2$, to the smallest domain size, $50 \times 50 \text{m}^2$, ranges from $40\%$ to $33\%$, due to a randomly distributed sea ice layout. %A constant FVM-cell size of $0.33 \times 0.33 \text{m}^2$
%$3.125 \times 3.125 \text{m}^2$ 
%is used in the simulations for this analysis. %Note the chosen cell size is larger than the optimal found in the previous section due to computer memory limitations encountered.
\\
\\
Figure~\ref{fig4} shows the overall domain-averaged sea ice rheology variables (stress, strain rate and bulk viscosity) in both ice floes and grease ice for different domain sizes and three different wave periods. Clearly, smaller wave periods reduce the required domain size for convergence. For the largest wave period of $T = 20 \text{s}$, the threshold is identified at an inner domain size of $1600 \times 1600 \text{m}^2$, where the blue and red curves in Figures~\ref{fig4}(c), (f) and (i), which represent an inner domain size of $1600 \times 1600 \text{m}^2$ and $3200 \times 3200 \text{m}^2$, respectively, converge. In contrast, for the smallest wave periods $T = 6 \text{s}$, the minimum required inner domain size is only $400 \times 400 \text{m}^2$.

\section{Analysis of the sea ice layout and rheology} \label{SA}

In this section the sea ice rheology variables are separately analysed for floes and grease ice. Section~\ref{DCA} indicates that any inner domain size smaller than $400 \times 400 \text{m}^2$ results in a different mechanical response for any wave period larger than $T = 6 \text{s}$. An inner domain size of $100 \times 100 \text{m}^2$ is therefore expected to produce temporal and spatial fluctuations of the strain rate distribution depending on the ratio of median pancake floe diameter and wavelength, $D_{p,median}/\lambda$, and thus, warrant detailed small-scale modelling. %For this reason, the threshold of this ratio is to be determined where the fluctuations become significant with respect to the temporally and spatially-averaged strain rate distribution. 

\begin{figure}[htb!]
\centerline{
\includegraphics[width=13cm]{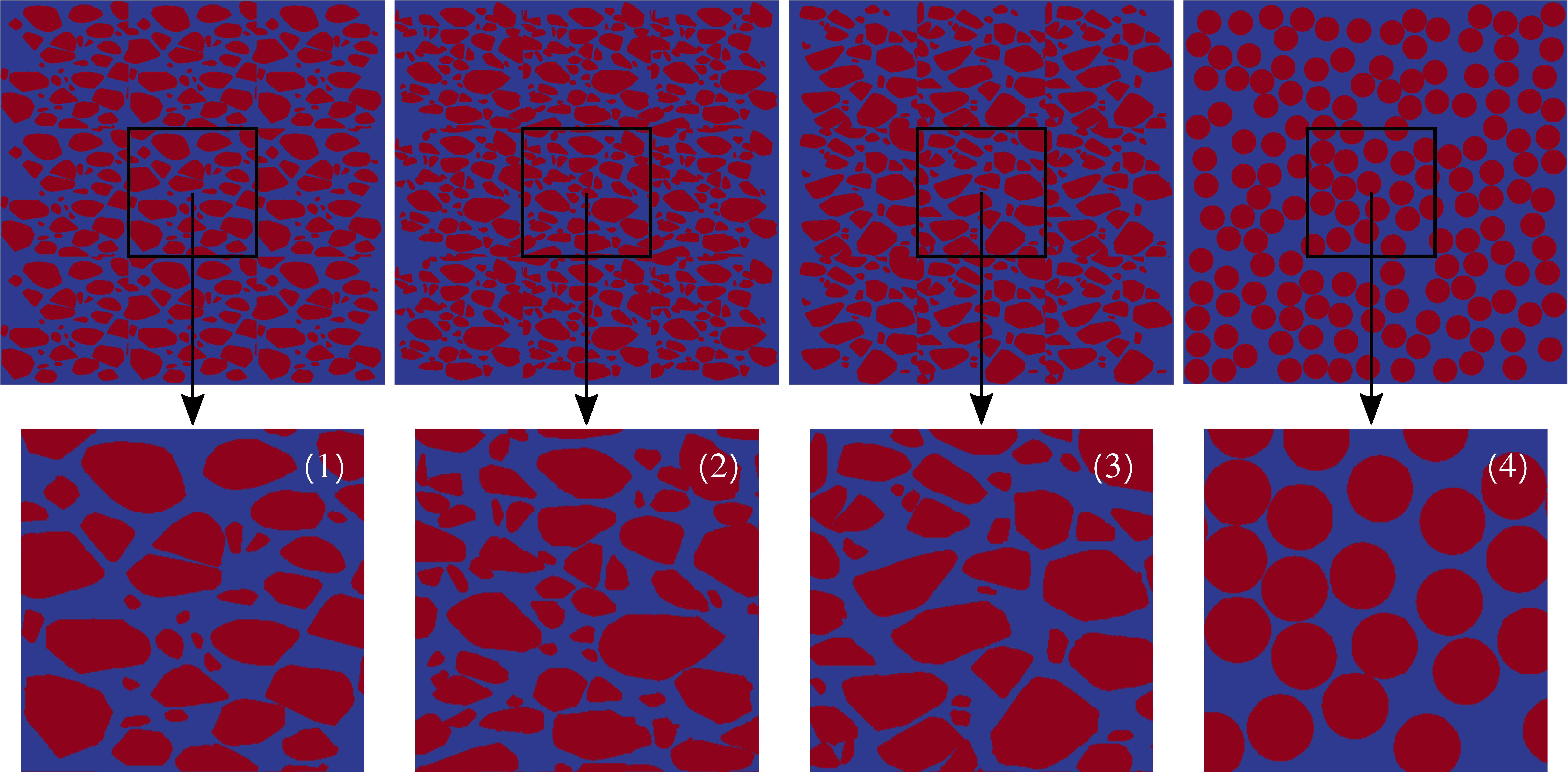}
}
	\caption{Three realistic sea ice layouts (1-3) and one idealised with disk-shaped floes (4), each with a $100 \times 100 \text{m}^2$ inner domain embedded in a $300 \times 300 \text{m}^2$ outer domain.}
\label{Fig3}
\end{figure}

Three realistic layouts and one idealised sea ice configuration comparable in terms of sea ice concentration are illustrated in Figure~\ref{Fig3}.
The ice floe caliper diameter ($D_x$ and $D_y$; in \textit{x}- and \textit{y}-direction), standard deviation (${SD}_x$ and ${SD}_y$; in \textit{x}- and \textit{y}-direction), ice floe concentrations, wave characteristics and viscosity are summarized in Table~\ref{tab2} for all four layouts.

\begin{table}[h] 
\captionsetup{justification=centering}
\begin{center}
\caption{Parameters used in the analysis of sea ice composition and rheology, where XCD and YCD denote the ice floe caliper diameter in \textit{x}- and \textit{y}-directions, respectively. The standard deviation in \textit{x}- and \textit{y}-directions is indicated by XSD and YSD, respectively.
\label{tab2}}
\begin{tabular}{l l l l}
\hline
\textbf{Parameter} & \textbf{Definition} & \textbf{Value} & \textbf{Unit} \\
\hline
 $D_{x,median}$ & median XCD layout 1, 2, 3, 4 & 13.0, 11.0, 9.3, 9.7 & $\text{m}$ \\
 $D_{y,median}$ & median YCD layout 1, 2, 3, 4 & 10.0, 7.0, 8.0, 9.7 & $\text{m}$ \\
  ${SD}_x$ & XSD layout 1, 2, 3, 4 & 8.8, 7.3, 9.2, 0 & $\text{m}$ \\
 ${SD}_y$ & YSD layout 1, 2, 3, 4 & 5.5, 4.3, 5.6, 0 & $\text{m}$ \\
 $A$ & ice floe concentration layout 1, 2, 3, 4 & 54.7, 59.7, 57.3, 59.7 & $\%$ \\
 $T$ & wave period & 8, 12, 16 & $\text{s}$ \\
 $a$ & wave amplitude & 1, 2.1, 3.8 & $\text{m}$ \\ 
 $\lambda$ & wavelength & 100, 225, 400 & $\text{m}$ \\ 
$\nu$ & grease ice viscosity & 0.04 & $\text{m}^2\text{s}^{-1}$ \\ 
$P^*$ & grease ice strength  & 0.024 & $\text{N}\text{m}^{-3}$  \\
 \hline
\end{tabular}
\end{center}
\end{table}

\begin{figure}[htb!]
\centerline{
\includegraphics[width=12cm]{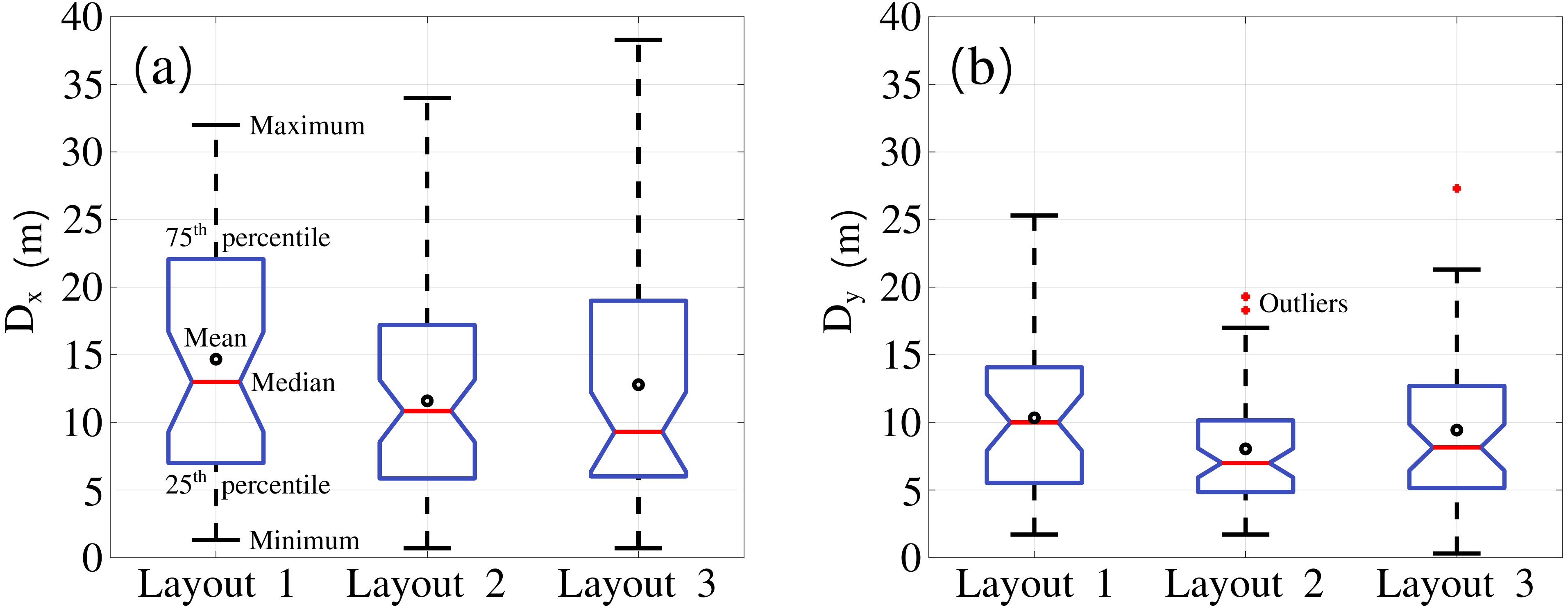}}
	\caption{Boxplots for all three realistic sea ice layouts, showing the distribution of the (a) caliper diameter in \textit{x}-direction and (b) caliper diameter in \textit{y}-direction.}
\label{boxplots}
\end{figure}

Layout differences in ice floe concentration, $57.9\%$ ($\pm 3.2\%$), and median ice floe caliper diameter with $10.8\text{m}$ ($\pm 2.2\text{m}$) in \textit{x}-direction and $8.7\text{m}$ ($\pm 1.7\text{m}$) in \textit{y}-direction, respectively, are small. The distribution of the ice floe caliper diameters has been mapped for all realistic sea ice layouts through the use of a boxplot, as shown in Figure~\ref{boxplots}. Layout 2 has the smallest interquartile range. %, i.e. the distance between the bottom ($25^{\text{th}}$ percentile) and top ($75^{\text{th}}$ percentile) of the boxplot. 
This is also reflected in the standard deviation, which show the lowest values in both \textit{x}- and \textit{y}-direction for layout 2. This indicates that it features the largest portion of medium-size floes and is the most homogeneous in size. Layout 3, on the other hand, exhibits the largest spread of floe sizes, in particular with regards to large floes. 

As layout 2 is the most homogeneous, its mechanical response is expected to be closest to the idealised sea ice layout 4 and is therefore chosen to study the error introduced by completely disregarding floe shape and variations of floe diameter. % Therefore, the ice floe concentrations in layout 2 and 4 are the same.
In order to specifically focus on the sensitivity regarding those two sea ice characteristics and study their effect on stress, strain rate and viscosity variables in both the grease ice and ice floe rheology, the ice floe concentration is chosen to be the same and the mean ice floe caliper diameter in layout 4 has been derived from layout 2, such as that the average area per ice floe, $A_{floe} = 293\text{m}^2$, is identical.

All sea ice layouts in this analysis are subjected to three different wave forcings via Equation~\ref{eq:1111} using wave periods ranging between $T =8-16\text{s}$ and for constant wave steepness $ak = 0.06$ (by varying the wave amplitude). The domain-averaged grease ice viscosity value is $\nu \approx 0.04 \text{m}^2 \text{s}^{-1}$ in agreement with literature values \cite{wadhamsa2006wave, newyear1999comparison, wang2010experimental}. The viscosity of the grease ice rheology, Equation~\eqref{eq:1313}, is strain rate-dependent via the ice strength parameter, $P$, which in turn depends on the empirical constant, $P^*$. The value of $P^*$ is chosen such as that the domain-averaged viscosity provides a close match to the predefined values above. For each sea ice layout, three simulations of $30 \text{s}$ are carried out according to the three aforementioned wave periods to elucidate the impact of wave length on the mechanical response.

%\begin{specialtable}[H] 
%\captionsetup{justification=centering}
%\begin{center}
%\setlength{\tabcolsep}{10pt}
%\setlength{\extrarowheight}{4pt}
%\caption{Overview of all 15 simulation variations per layout used for the sensitivity analysis. \label{tab3}}
%\begin{tabular}{|>{\centering\arraybackslash}m{0.2in}|c|c|c|c|}
%\hline
%\multicolumn{2}{|c|}{} & \multicolumn{3}{c|}{\textbf{Kinematic viscosity} $\text{m}^2 \text{s}^{-1}$} \\ \cline{3-5}
%\multicolumn{2}{|c|}{} & $\nu \approx 0.16$ & $\nu \approx 0.04$ & $\nu \approx 0.01$ \\
%\hline
%\multirow{5}{*}{\rotatebox[origin=p]{90}{\textbf{Wave properties}}} & $T = 20\text{s} \hspace{1mm} (\Lambda = 625 \text{m}), a = 6.0\text{m}$ & $T_{20},\nu_{0.16}$ & $T_{20},\nu_{0.04}$ & $T_{20},\nu_{0.01}$ \\ \cline{2-5} 
%& $T = 18\text{s} \hspace{1mm} (\Lambda = 506 \text{m}), a = 4.8\text{m}$ & $T_{18},\nu_{0.16}$ & $T_{18},\nu_{0.04}$ & $T_{18},\nu_{0.01}$ \\ \cline{2-5} 
%& $T = 16\text{s} \hspace{1mm} (\Lambda = 400 \text{m}), a = 3.8\text{m}$ & $T_{16},\nu_{0.16}$ & $T_{16},\nu_{0.04}$ & $T_{16},\nu_{0.01}$ \\ \cline{2-5} 
%& $T = 12\text{s} \hspace{1mm} (\Lambda = 225 \text{m}), a = 2.1\text{m}$ & $T_{12},\nu_{0.16}$ & $T_{12},\nu_{0.04}$ & $T_{12},\nu_{0.01}$ \\ \cline{2-5} 
%& $T = 8\text{s} \hspace{1mm} (\Lambda = 100 \text{m}),  a = 1.0\text{m}$ & $T_{8},\nu_{0.16}$ & $T_{8},\nu_{0.04}$ & $T_{8},\nu_{0.01}$  \\
%\hline
%\end{tabular}
%\end{center}
%\end{specialtable}

Figure~\ref{fig:layout 1,2,3,4} shows the spatially-averaged stress and strain rate magnitudes in both grease ice and pancake ice floes for wave periods $T = 8\text{s}$, $T = 12\text{s}$ and $T = 16\text{s}$ with $\nu \approx 0.04 \text{m}^2 \text{s}^{-1}$. In Figure~\ref{fig:layout 1,2,3,4}(a-c) the spatially-averaged ice floe stress magnitude evolution over time is shown for all sea ice layouts. %The sea ice concentration and mean ice floe diameter are comparable for all sea ice layouts, see Table~\ref{tab2}.
The average stress in ice floes is similar for all wave periods due to a prescribed wave steepness, $ak=0.06$. An increasing wave period results in an increasing stress amplitude and a decreasing stress frequency. The discrepancy between the stress curves of the four considered layouts increases for a decreasing wave period signifying the mounting influence of the detailed heterogeneous sea dynamics description. Clearly, the realistic sea ice layout 1 exhibits a smaller floe stress magnitude compared to layouts 2 and 3, in particular for the smallest wave period, $T = 8\text{s}$, as illustrated in Figure~\ref{fig:layout 1,2,3,4}(a). Additionally, it can be observed that the floe stress differences between layouts change with the wave period and that the grease ice stress fluctuations, shown in Figure~\ref{fig:layout 1,2,3,4}(d-f), in particular for $T = 8\text{s}$, are considerably lower than seen for the floe stress. %This seems to imply that the Froude-Krylov force is the main source for those observations considering that it solely acts on the ice floe circumference. As such, the dependency on the ratio of floe diameter to wave length becomes significant. It might be also plausible that the relatively large amount of larger floe diameters in layout 1 reduces the wave-floe interaction and thus, floe collision due to larger inertia. Both hypotheses, however, warrant further in-depth investigations.
With regards to influence of floe shape and diameter variations, the floe stress and strain rate response of the idealized sea ice composition (layout 4) is distinctly different from the realistic layout 2, both being underestimated by the idealization of ice floe geometry. The average discrepancy in the ice floe stress curves between layout 2 and 4 is approximately $7\%$.

\begin{figure}[htb!]
\centerline{
\includegraphics[width=13.5cm]{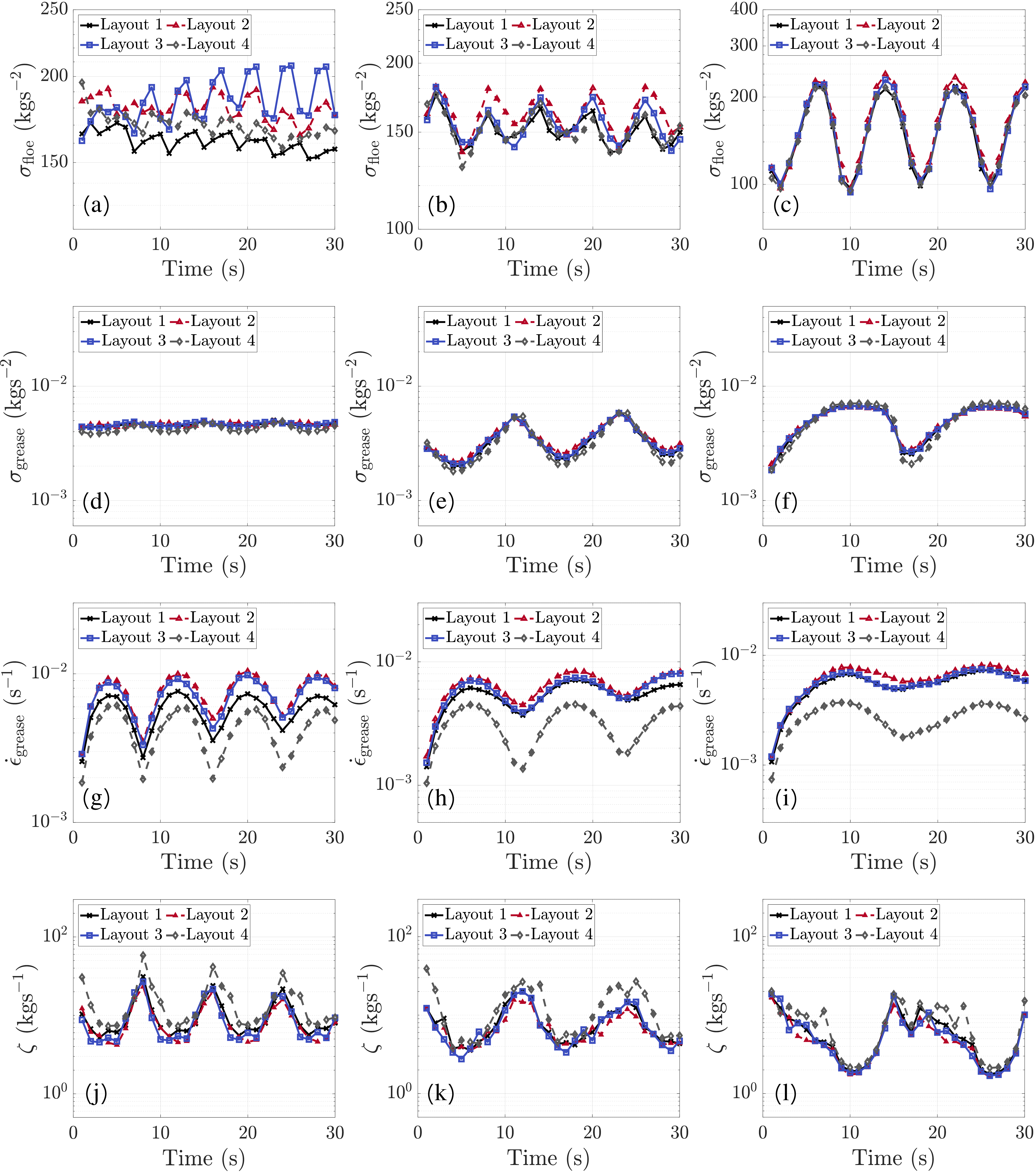}
}
	\caption{Spatially-averaged mechanical sea ice response for layouts 1-4 with $\nu \approx 0.04 \text{m}^2 \text{s}^{-1}$ showing ice floe stress for $T = 8\text{s}$ (a), $T = 12\text{s}$ (b), $T = 16\text{s}$ (c); grease ice stress for $T = 8\text{s}$ (d), $T = 12\text{s}$ (e), $T = 16\text{s}$ (f); grease ice strain rate for $T = 8\text{s}$ (g), $T = 12\text{s}$ (h), $T = 16\text{s}$ (i); bulk viscosity for $T = 8\text{s}$ (j), $T = 12\text{s}$ (k), $T = 16\text{s}$ (l).}
\label{fig:layout 1,2,3,4}
\end{figure}

As mentioned previously, Figure~\ref{fig:layout 1,2,3,4}(d-f) show the spatially-averaged grease ice stress for sea ice layouts 1-4. Results of the three realistic sea ice layouts (layout 1-3) are very similar for all wave periods, due to a comparable concentration of grease ice in all layouts. On the other hand, the uniformity of floe diameter and shape results in an average discrepancy in the grease ice stress between layout 2 and the idealized one (layout 4) of about $7\%$.

Lastly, Figure~\ref{fig:layout 1,2,3,4}(g-l) show the spatially-averaged grease ice strain rate and bulk viscosity for all sea ice layouts. As for the flow stress, the discrepancy between the strain rate evolution in time increases for decreasing wave period looking at the realistic sea ice layouts and a distinctly smaller strain rate magnitude is exhibited for layout 1. Comparing the grease ice strain rate curves of layout 2 and 4, we clearly see that the average strain rate is lower for layout 4, as it was the case for floe and grease ice stress. The average discrepancy in the grease ice strain rate between layout 2 and 4 is substantial with approximately $103\%$. From Equation~\eqref{eq:8} we know that the relation between the grease ice strain rate and viscosity is inversely proportional. An increase in grease ice viscosity, results in a decreasing strain rate and vice versa. As these variables are directly related, the average discrepancy in grease ice viscosity between layout 2 and 4 is also quite high, with a value of roughly $42\%$. 
%\begin{figure}[htb!]
%\centerline{
%\includegraphics[width=13.5cm]{stress-strainrate-viscosity-L2and4.pdf}
%}
%	\caption{Spatially- and temporally-averaged magnitudes for all simulations of (a) grease ice stress versus viscosity for layout 2 (b) grease ice strain rate versus viscosity for layout 2 (c) grease ice stress versus viscosity for layout 3 (d) grease ice stress versus strain rate for layout 1 (e) grease ice stress versus strain rate for layout 2 (f) grease ice stress versus strain rate for layout 3 (g) grease ice strain rate versus viscosity for layout 1 (h) grease ice strain rate versus viscosity for layout 2 (i) grease ice strain rate versus viscosity for layout 3.}
%\label{fig:sensitivity grease stress layout 1,2,3}
%\end{figure}
%%%%%%%%%%%%%%%%%%%%%%%%%%%%%%%%%%%%%%%%%%
\section{Conclusions} \label{D}

In this paper, we presented a new numerical approach to model the small-scale interaction of waves, pancake ice floes and interstitial grease ice. %the phenomenological model by \citet{Hibler1979} has been modified on several aspects. 
For this, the material behaviour of ice floes and grease ice is separately described considering for both characteristic material properties. Ice floes are modelled as elastic solids with high stiffness using generalised Hooke's law. Grease ice behaves as a viscous fluid obeying a viscous-plastic flow rheology. This two-phasic description is in contrast to the commonly used smeared phenomenological approaches in which a large heterogeneous sea ice area is homogenised as one isotropic, continuous material with averaged quantities.
To study the influence of wave forcing on the mechanical response of sea ice, a linearised harmonic propagating wave has been imposed on the domain via the Froude-Krylov force emulating the wave pressure field acting on the ice floe circumference and the skin drag acting on the ice floe basal plane. The form drag on the ice floe due to grease ice is implicitly accounted for by the continuum approach comprising both ice constituents.

%This paper describes the computational fluid dynamics model set-up by showing a grid size and domain size convergence analysis. As the dynamics depends on a variety on ice type and characteristics, applied material law and wave forcing, a careful analysis of their impact on the mechanical response as linked to stress, strain rate and viscosity is performed. 
To justify small-scale modelling a domain-size threshold was identified as smaller than $400 \times 400 \text{m}^2$ where temporal and spatial fluctuations of the sea ice rheology variables (stress, strain rate and viscosity) become significant for all considered wave periods $T = \SIrange{6}{20}{\s}$ with $20\text{m}$-diameter ice floes. This wave period-dependent threshold marks the transition from small- to large-scale modelling where a phenomenological model with homogenized material properties can be utilized.
%This allows us to study the detailed mechanical response of the sea ice rheology on smaller scale, which is different for any domain size smaller than the threshold.

Simulating three realistic sea ice layouts of $100 \times 100 \text{m}^2$ extracted from in-situ images of the Antarctic MIZ, the robustness of the approach has been demonstrated. Furthermore, it was found that mechanical sea ice response tends to become independent of the detailed distribution of floes for wave periods larger than $T=16\text{s}$ in case of homogeneous sea ice conditions with similar ice concentrations and median floe caliper diameters.
%($10.8\text{m}$ ($\pm 12\%$) in \textit{x}-direction and $8.7\text{m}$ ($\pm 14\%$) in \textit{y}-direction). 
Moreover, the discrepancy between stress and strain rate curves increases for smaller wave period. An increasing ratio of ice floe diameter to wavelength results in a reduced stress and strain rate response in the considered realistic sea ice layouts. Based on the observation of significantly larger differences in floe stress between layouts compared to that of grease ice, it seems that the Froude-Krylov force is the main source for those observations considering that it solely acts on the ice floe circumference. As such, the dependency on the ratio of floe diameter to wave length becomes significant. It might be also plausible that the relatively large amount of larger floe diameters in layout 1 reduces the wave-floe interaction and thus, floe collision due to larger inertia. Both hypotheses, however, warrant further in-depth investigations.

Lastly, the influence of floe shape and diameter variations was shown to be significant considering additionally an idealized sea ice layout with disk-shaped floes with identical ice concentration and average area per ice floe. The domain-averaged strain rate magnitude and bulk viscosity are mostly affected by the detailed geometrical floe properties, with an average discrepancy of $103\%$ and $42\%$, respectively.

In summary, results of this paper demonstrate the importance of detailed small-scale modelling on the sub-kilometer scale to resolve the mechanical response of a heterogeneous sea ice cover due to the complex interaction of waves, floes and grease ice.
%This work has shown the initial small-scale investigation to wave-ice floe interaction embedded in grease ice.
%More research will follow to get a more in-depth understanding of the small-scale behaviour of both ice floes and grease ice. 
If large-scale regional models are to be informed by the actual material behaviour as originated on smaller scale, averaged material parameters need to be obtained from homogenization and up-scaling procedures. These rely on the identification of the minimum domain size threshold where the actual kinematics and material composition of the problem must be addressed in detail and the averaged quantities are statistically representative for larger domains.

%A sensitivity analysis is conducted where realistic sea ice layouts are subjected to different wave characteristics and grease ice viscosity values. 
%Note that in this paper the realistic sea ice layouts obtained from in-situ image and video material recorded in the Antarctic MIZ are projected on a $100 \text{x} 100 \text{m}^2$ domain size, although the actual domain size from the recorded material is smaller than $100 \text{x} 100 \text{m}^2$.
%The domain-averaged ice floe stress magnitude is similar for all wave periods, for a constant  wave steepness, $ka = 0.06$. An increasing wave period results in an increasing stress amplitude and a decreasing stress frequency. %By comparing three realistic sea ice layouts, it was found that the sea ice composition does not significantly affect the mean stress and strain rate magnitude distribution in the domain, as all three layouts have similar sea ice concentrations and mean floe caliper diameters.

\section*{Acknowledgement}
This research has been supported by the National Research Foundation of South Africa (Grant Numbers 104839 and 105858). Opinions expressed and conclusions arrived at, are those of the author and are not necessarily to be attributed to the NRF.
Computations were performed using facilities provided by the University of Cape Town's ICTS High Performance Computing team: \url{http://hpc.uct.ac.za}.
%This research has been supported by the Centre for High Performance Computing South Africa.

\section*{References}
%\begin{center}
%{\bf REFERENCES}\\[-3mm]
%\end{center}

\begingroup
\bibliographystyle{plainnat}
\renewcommand{\section}[2]{}
\bibliography{references}
\endgroup
\end{document}